%% file: LIV_Mrk421-2014_forJCAP_v5.tex
\documentclass[a4paper,11pt]{article}
\pdfoutput=1 

\usepackage{jcappub} 

\usepackage[T1]{fontenc} 

\usepackage{glossaries}

\include{commands}

\usepackage{makecell}
\usepackage{xcolor}
\usepackage{ulem}

\title{Constraints on Lorentz invariance violation from the extraordinary Mrk\,421 flare of 2014 using a novel analysis method}

\include{opted-in_collaborators_jcap_poll_2177_date_2023-11-29}

\emailAdd{contact.magic@mpp.mpg.de}

\abstract{The Lorentz Invariance Violation (LIV), a proposed consequence of certain quantum gravity (QG) scenarios, could instigate an energy-dependent group velocity for ultra-relativistic particles. This energy dependence, although suppressed by the massive QG energy scale $E_\mathrm{QG}$, expected to be on the level of the Planck energy $1.22 \times 10^{19}$ GeV, is potentially detectable in astrophysical observations. In this scenario, the cosmological distances traversed by photons act as an amplifier for this effect. By leveraging the observation of a remarkable flare from the blazar Mrk\,421, recorded at energies above 100 GeV by the MAGIC telescopes on the night of April 25 to 26, 2014, we look for time delays scaling linearly and quadratically with the photon energies. Using for the first time in LIV studies a binned-likelihood approach we set constraints on the QG energy scale. For the linear scenario, we set $95\%$ lower limits $E_\mathrm{QG}>2.7\times10^{17}$\,GeV for the subluminal case and $E_\mathrm{QG}> 3.6 \times10^{17}$\,GeV for the superluminal case. For the quadratic scenario, the  $95\%$ lower limits for the subluminal and superluminal cases are $E_\mathrm{QG}>2.6 \times10^{10}$\,GeV and $E_\mathrm{QG}>2.5\times10^{10}$\,GeV, respectively. }

\keywords{Gamma rays, Lorentz Invariance Violation, Time-of-Flight analysis, Active Galactic Nuclei, MAGIC telescopes, statistical data analysis}

\begin{document}
\maketitle
\flushbottom

\section{\label{sec:Introduction}Introduction}

The exploration of quantum gravity (QG) --- a research field dedicated to describing the quantum behavior of the gravitational field --- has arisen from the fundamental incompatibility issues \cite{UVGravity,QuantizingGrav,QGTheories,QuantizationFields, ExampleGravCollapseGR} between general relativity and quantum field theory, the two central pillars of modern physics.
Numerous QG theoretical frameworks, such as string theory \cite{LIVStringTheory}, space-time foam \cite{QGtestsGRB}, loop quantum gravity \cite{OpticsQG}, non-commutative geometry \cite{NonCommutativeLIV, kappaSystems, WavesNonComm}, and brane-world backgrounds \cite{LoopGraviton}, allow for distinct violations of Lorentz symmetry, giving rise to the concept of Lorentz Invariance Violation (LIV).
LIV can be encapsulated through the addition of gauge invariant and renormalizable terms to the Standard Model Lagrangian, violating Lorentz invariance, in what is known as the Standard Model Extension (SME) approach \cite{LIVStringTheory}.
Alternatively, the symmetry deformation approach known as double special relativity (DSR) upholds Lorentz invariance but modifies the transformation laws \cite{RelativityShortDistance, DSRfacts}.  For a comprehensive comparison and details regarding different approaches, please refer to \cite{ModernTestsLI, QSphenomenology, Leyrethesis, Addazi:2021xuf}.

The energy scale of QG, where QG effects should manifest, is expected to be on the level of the Planck energy $1.22 \times 10^{19}$\,GeV. 
Assuming a massless particle and considering that the typical energy $E$ of an observable gamma-ray is significantly lower than the QG energy scale ($E \ll E_\mathrm{QG}$), the photon dispersion relation can be represented as a Taylor expansion as follows:

\begin{equation}\label{eq:moddispastro}
E^{2}=p^{2} c^{2} \left[1+\sum_{n=1}^{\infty} S_{\mathrm{n}}\left(\frac{E}{E_{\mathrm{QG}, n}}\right)^{n}\right].
\end{equation}
In this equation, $E$ and $p$ are the energy and momentum of a very high energy (VHE, E > 100\,GeV) gamma ray, respectively, while $c$  denotes the Lorentz invariant speed of light. $E_{\mathrm{QG}, n}$ are the energy scales of QG effects. $n$ represents the order of a correction to the dispersion relation. The parameter $S_{\mathrm{n}}$ can have values of either $+1$ or $-1$.



Astrophysical testing through Time of Flight (ToF) measurements of photons from distant sources is a common method for scrutinizing modified dispersion relations, offering significant constraints on $E_{\mathrm{QG},n}$ \cite{QGtestsGRB, AstroProbesVelocityLight, QGAanalysisWavelets}. 
From Equation~ \ref{eq:moddispastro} we can determine the adjusted group velocity of a photon in vacuum as:

\begin{equation}\label{eq:photonvelocity}
v_{\gamma}=\frac{\partial E}{\partial p} \simeq c\left[1+\sum_{n=1}^{\infty} S_{n} \frac{n+1}{2}\left(\frac{E}{E_{\mathrm{QG}, n}}\right)^{n}\right].
\end{equation}
The implication of an energy-dependent group velocity is that photons of different energies, emitted simultaneously from the source, will reach the detector at different times. 
Considering a photon with energy $E$, the delay $\Delta t$ with respect to its arrival time in case no LIV effect were present   can be expressed as \cite{DistancePiron}: 

\begin{equation}\label{eq:delay}
    \Delta t_n \cong - S_{\mathrm{n}}  \frac{n+1}{2} \frac{E^{n}}{ E_{\mathrm{QG}, n}^{n}} \kappa_n (z_{\mathrm{s}}) \; \equiv  \eta_n   E^n,
\end{equation}
where the parameters $S_{\mathrm{n}}$ have value  $-1$ in the scenario where higher-energy photons arrive later (subluminal) or ($+1$) in the scenario where they arrive earlier (superluminal). The $\kappa_n (z_{\mathrm{s}})$ is a parameter dependent on the distance of the source (with $z_{\mathrm{s}}$ the source's redshift), and it plays the role of an amplifier of the process, compensating for the small ratio $E/E_{\mathrm{QG}, n}$. 
Experimental LIV searches usually use the form first derived in~\cite{DistancePiron}:

\begin{equation}\label{eq:distanceToFPiran}
    \kappa_n (z_{\mathrm{s}}) = \frac{1}{H_0}\int_0^{z_{\mathrm{s}} }\frac{(1+z)^n}{\sqrt{\Omega_\Lambda + \Omega_m(1+z)^3}}\;\mathrm{d}z,
\end{equation}
where $\Omega_\Lambda \cong 0.69$ and $\Omega_m \cong 0.31$ are the standard cosmological parameters of the $\Lambda$CDM cosmology model \cite{CosmoParameters} and $H_0 \equiv 70$ km  Mpc$^{-1}$ s$^{-1}$ is the Hubble constant (the impact of selecting different values for the Hubble constant on our analysis is examined in appendix \ref{Systematics}). In Equation (\ref{eq:delay}) we have introduced the ``spectral lag'' parameter which has the dimension of time over the $n$-th power of energy, and it is related~\footnote{This relationship between the spectral lag parameter and the QG energy scale presupposes the absence of intrinsic spectral lag \cite{perennes2020modeling} within the observed data. In other words, it assumes that the flare is intrinsically achromatic and that any observed correlation between the energy and the photon's time of arrival is exclusively attributable to LIV effects.  } to the QG energy scale through:
\begin{equation}
     \eta_n = -  S_{\mathrm{n}} \frac{n+1}{2} \frac{\kappa_n (z_{\mathrm{s}})}{E_{\mathrm{QG}, n}^n}.
     \label{eq:spectral_lag}
\end{equation}

A potential lack of a significant ToF difference between photons of different energies will enable us to set upper limits (ULs) on $\eta_n$, that can be translated to lower limits (LLs) on $E_{\mathrm{QG}, n}$. In this study, we will concentrate exclusively on the linear ($n=1$) and quadratic ($n=2$) cases, as these are the only scenarios we can probe through gamma-ray astronomical observations~\footnote{For $ n > 2 $, achieving observable time delays of one second or greater necessitates allowing the quantum gravity energy scale $ E_{\mathrm{QG}} $ to be on the order of the PeV scale, or even lower.
 }.

In gamma-ray astronomy, sources suitable for LIV investigations using ToF must exhibit very-high-energy emissions, be located at considerable distances, and display rapid variability. Gamma-ray detectors have hitherto utilized three types of sources for LIV studies: Gamma-Ray Bursts (GRBs), Pulsars, and Active Galactic Nuclei (AGNs) flares. 
%
Some of the most stringent constraints, outlined in Table \ref{tab:ArchiveLIVResults}, have been acquired from observations performed with the \textit{Fermi}Large Area Telescope (LAT) \cite{Limit4GRBFermi}, the Major Atmospheric Gamma-ray Imaging Cherenkov (MAGIC) telescopes \cite{magic2019teraelectronvolt,acciari2020bounds,LimitMAGICMrk501,LimitMAGICCrab}, the High Energy Stereoscopic System  (H.E.S.S.) telescopes  \cite{CedricPaper,LimitHESSPKS2155Like}, or, very recently, from the Large High Altitude Air Shower Observatory (LHAASO) \cite{piran2023lorentz}. 
The constraints on the QG energy scale in the linear scenario have already surpassed the Planck scale, which is where QG effects should manifest. However, the quadratic scenario constraints are still significantly below, implying potential for further advancements by ground-based gamma-ray instruments, given that energy is a dominant parameter in this scenario (see, e.g., Table 1 in~\cite{Terzic:2021rlx}).

\begin{table*}
	\centering
\begin{tabular}{|ccccccc|}
		\hline
		{\bf Source} & {\bf Source} & {\bf Redshift/} & $\mathbf{E_\mathrm{QG,1}}$ & $\mathbf{E_\mathrm{QG,2}}$ & {\bf Instrument} & {\bf Ref.}\\
		{\bf name} & {\bf type}  & {\bf Distance} & $\mathbf{10^{19}}$\, {\bf GeV} & $\mathbf{10^{10}}$\, {\bf GeV} &  & \\
		\hline
	    GRB\,090510 & GRB & 0.9 &\makecell{($-$) $2.2$ \\ (+) $3.9$}&  \makecell{($-$) $4.0$ \\ (+) $3.0$}& \textit{Fermi}-LAT & \cite{Limit4GRBFermi}\\ \hline
     GRB\,190114C & GRB & 0.4245 &\makecell{($-$) $0.58$ \\ (+) $0.55$} & \makecell{($-$) $6.3$ \\ (+) $5.6$}  & MAGIC & \cite{acciari2020bounds}\\ \hline
     GRB\,221009A & GRB & 0.151 &\makecell{($-$) $7.2$ \\ (+) $7.6$} & \makecell{($-$) $71$ \\ (+) $56$}  & LHAASO & \cite{piran2023lorentz}\\ \hline
		Mrk501 & AGN & 0.034 & \makecell{($-$) $0.036$ \\ (+) $0.027$}& \makecell{($-$) $8.5$ \\ (+) $7.3$} & H.E.S.S. & \cite{CedricPaper}\\ \hline
		PKS 2155-304 & AGN & 0.116 &($-$) $0.21$ &($-$) $6.4$ & H.E.S.S. & \cite{LimitHESSPKS2155Like}\\ \hline
		Mrk501 & AGN & 0.034 &($-$) $0.021$ &($-$)  $2.6$ & MAGIC & \cite{LimitMAGICMrk501}\\ \hline
		Crab Pulsar & Pulsar & 2.0 kpc & \makecell{($-$) $0.055$ \\ (+) $0.045$} & \makecell{($-$) $5.9$ \\ (+) $5.3$} & MAGIC & \cite{LimitMAGICCrab}\\\hline
\end{tabular}
\label{tab:ArchiveLIVResults}
\caption{List of the most stringent LLs on $E_\mathrm{QG,1}$ and $E_\mathrm{QG,2}$ based on ToF studies. Markers (+) and ($-$)  represent superluminal and subluminal behaviours, respectively. The
LLs are expressed on the 95\% confidence level (CL).}
\end{table*}

In this study, we present ToF measurements derived from the 2014 flare of Mrk\,421, as observed by the MAGIC telescopes. This study also marks the first implementation of an innovative statistical method that does not require prior knowledge of the intrinsic temporal distribution of gamma rays. 
The structure of the paper is as follows: Section \ref{sec:DataSet} details the MAGIC data from the 2014 flare of Mrk\,421 used in this study. Section \ref{sec:LIV_Analysis} outlines the novel binned-likelihood method for LIV analysis, marking its first application in this area of research.   We present our findings in Section \ref{sec:Results} and the systematic uncertainties affecting our results in Appendix  \ref{sec:Systematics}. Finally, we discuss these findings in Section \ref{sec:Discussion}.

\section{\label{sec:DataSet}MAGIC observations and data analysis}

The MAGIC telescopes comprise two 17-meter diameter Imaging Atmospheric Cherenkov Telescopes (IACTs). Situated at the Roque de los Muchachos Observatory in La Palma, Canary Islands, Spain, at an altitude of $\sim 2200$ meters above sea level \cite{MAGICUpgradeSoftware}, these telescopes are specifically optimized for observing Cherenkov-light flashes generated by VHE gamma rays in the atmosphere \cite{MAGICUpgradeSoftware}. The telescopes typically operate in a stereoscopic mode, wherein only events simultaneously observed by both telescopes are recorded and analyzed \cite{MAGICUpgradeSoftware}. The standard trigger threshold for the MAGIC telescopes in low zenith angle observations under dark conditions, with high atmospheric transparency, is approximately 50 GeV \cite{MAGICUpgradeSoftware}. However, the threshold may increase under suboptimal atmospheric conditions, higher zenith angle and background light.

The AGN Mrk\,421, hosted by the galaxy UGC 6132 at a redshift of z=0.031 \cite{1975ApJ...198..261U}, is the first extragalactic object discovered in the VHE domain \cite{Mrk421detection} and remains one of the most intensively studied sources in this energy range to date.
During a regular monitoring observation on the night of April 25 to 26, 2014 Mrk\,421 exhibited an exceptional gamma-ray flare. The total observation lasted for approximately 3:40 hours. 
The flux at energies above 100 GeV reached approximately 8 Crab Units (CU\footnote{These units are defined relative to the observed electromagnetic radiation flux from the Crab Nebula. Since the Crab Nebula emits strongly across a wide range of the electromagnetic spectrum and its emission is approximately constant in time, it serves as a standard reference. For example, a 2 CU flux means that the flux is twice as much as the flux from the Crab Nebula in the same energy band. We use a value obtained in \cite{2015JHEAp...5...30A} as a reference value. }), representing an extraordinary 16-fold increase over Mrk 421's typical state of 0.5 CU \cite{acciari2014observation}.
All observations were conducted in the ``wobble'' mode \cite{fomin94, MAGICUpgradeSoftware} under dark sky conditions \cite{MAGICUpgradeHardware}, with aerosol transmission, as measured by the MAGIC LIDAR system \cite{fruck2022,2023A&A...673A...2S}, exceeding 90\% at 9\,km above the ground, throughout the observation period. 
The data underwent standard reduction and analysis using the MAGIC Analysis and Reconstruction Software (MARS) \cite{MARS, MAGICUpgradeSoftware}. 

In this study, we establish a reference time $T_0$ corresponding to the start of the first observational run considered in our analysis. All ensuing time measurements will be expressed relative to this reference point $T_0$, corresponding to April 25, 2014, at 22:26:34 Universal Time (UT).  We focused on the observations up to 35\,deg in zenith distance, resulting in a total of 9 runs (each run has an associated pointing sky coordinate and lasts approximately 15 minutes), which enabled us to set the analysis energy threshold at 100\,GeV.
After data quality cuts, we measured approximately $9 \times 10^3$ events above 100\,GeV in the signal region, with the last event being detected 8136 seconds after $T_0$ and with only about 7\% of them expected to be background contamination.
We have transformed the proprietary \texttt{melibea} files used by MAGIC, which contain reconstructed stereo event information, into the standardized Data Level 3 (DL3) format. The DL3 format is the standard adopted by the next-generation Cherenkov Telescope Array (CTA) consortium, as described in \cite{Nigro:2021xcr}. This format conversion facilitates the use of the open-source \texttt{gammapy} \cite{donath2023gammapy,acero_fabio_2023_7734804} software --- a cross-platform, multi-instrument tool for gamma-ray astronomy which is already widely employed in the analysis of existing gamma-ray instruments like H.E.S.S., MAGIC, VERITAS, and HAWC \cite{2019A&A...625A..10N}, and is poised to be the core library for CTA's Science Analysis tools. The list containing arrival times and reconstructed energies of individual events was used to perform the ToF analysis. Signal extraction in this study relies on one On region and three Off regions. The On region, also known as the Region Of Interest, is where the source's signal is expected. The number $N_{\mathrm{on}}$ of events detected in this region contains both the signal (potentially) and an irreducible number of background events. These background events are predominantly produced by protons, electrons, and cosmic ray nuclei.  To estimate this number, we used three Off, background-control regions, supposedly void of any signal and where $N_{\mathrm{off}}$ events are detected. 


\subsection{\label{sec:SED}Spectral energy distribution}

\begin{figure}[h]
    \centering
    \includegraphics[width=0.49\textwidth]{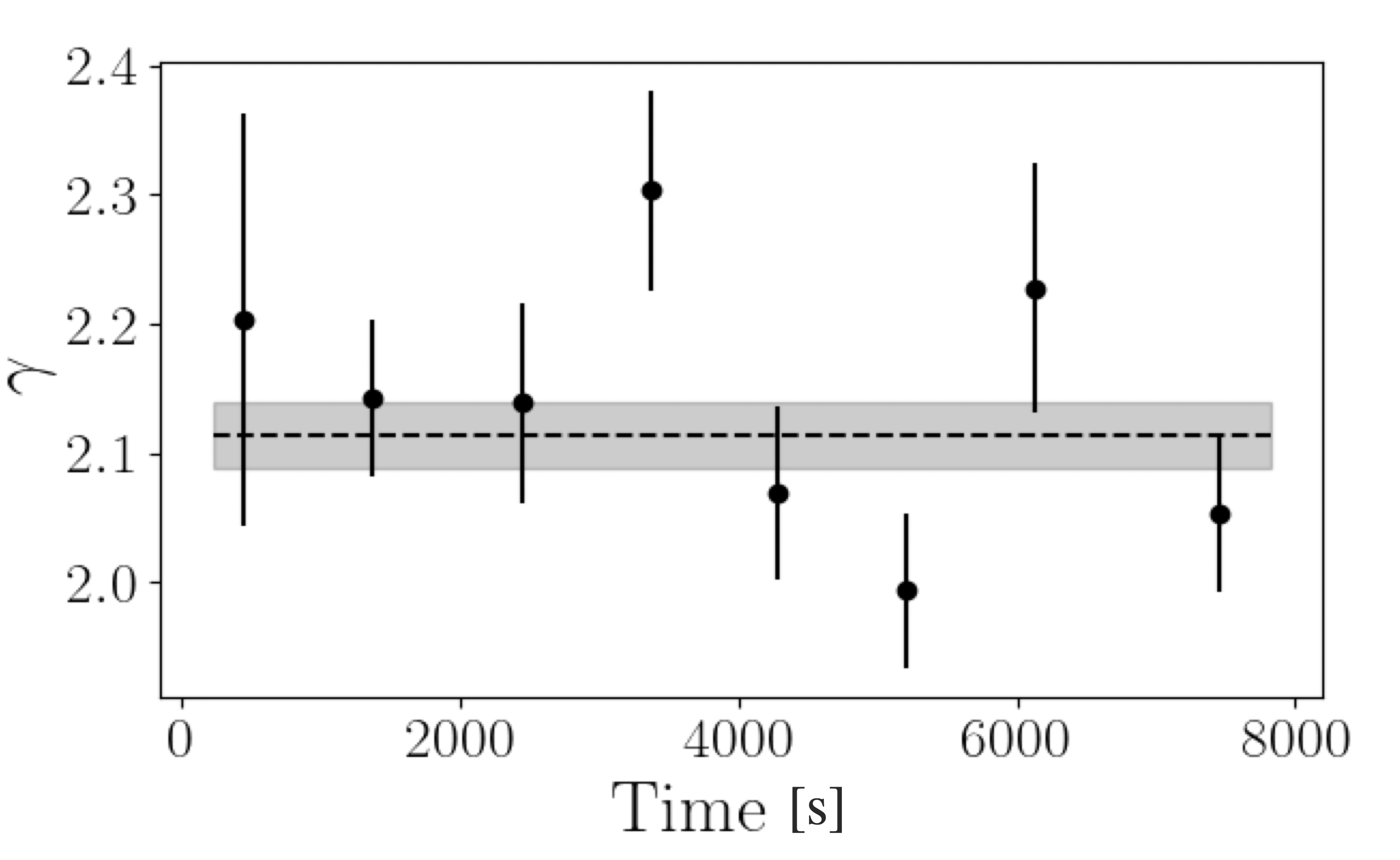}
    \includegraphics[width=0.49\textwidth]{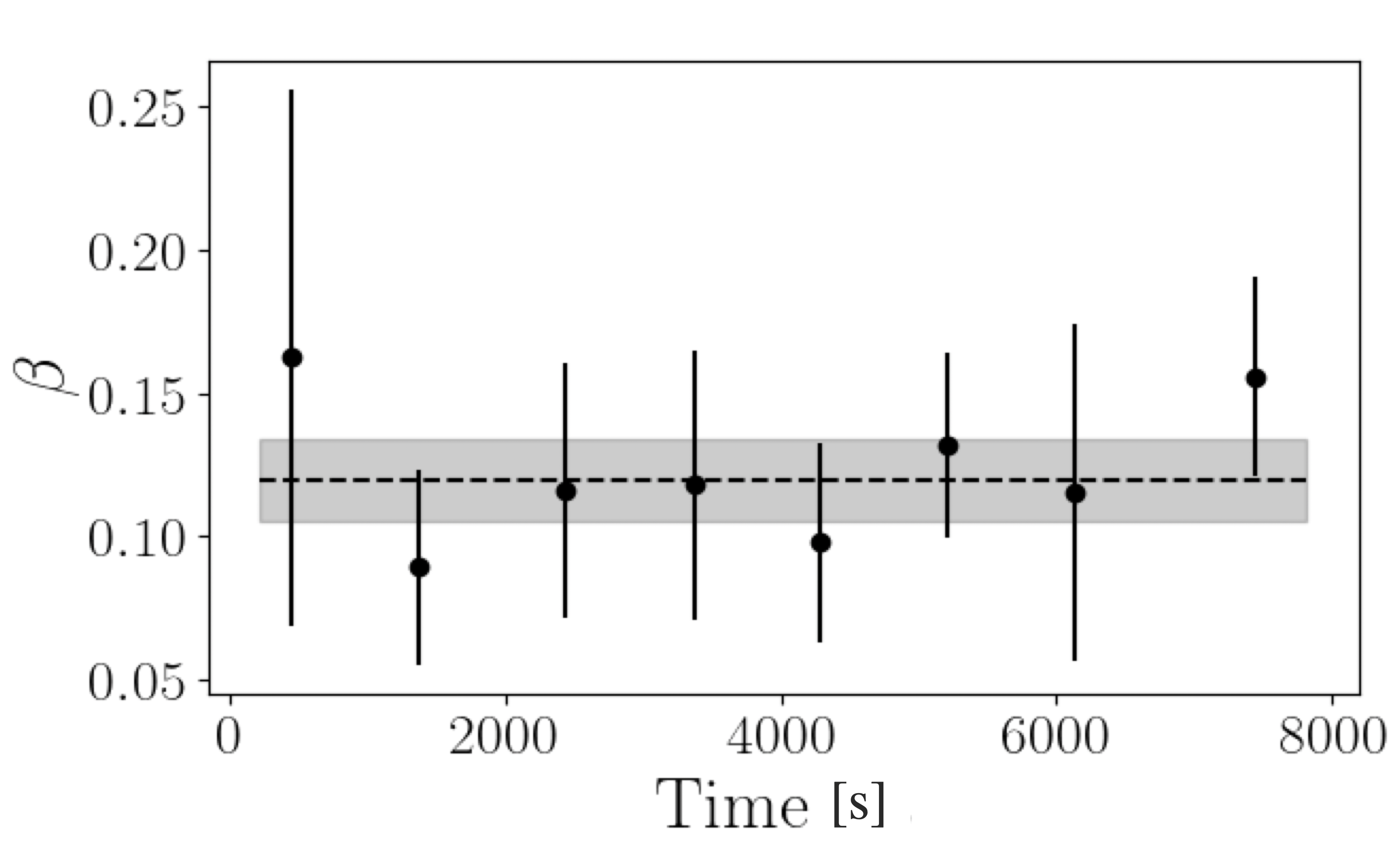}
    \caption{Temporal evolution of the SED parameters $\gamma$ (left panel) and $\beta$ (right panel) obtained from forward folding analysis on individual runs. The black dashed lines represent the best fit constant values, $\gamma = 2.11 \pm 0.03$ and $\beta = 0.119 \pm 0.014$, while their respective uncertainties are shown as grey bands. To enhance statistical robustness, the final two runs were combined. The consistency of these parameters over time indicates the stability of the source's spectral behavior during the observation.} 
    \label{fig:SED_paramters}
\end{figure}

The intrinsic energy distribution is  described ($\chi^2/\mathrm{dof} = 20.19/14$, p-value = 12.4\%) with a log-parabolic model given by

\begin{equation} \label{eq:SpectrumEquation}
\frac{d \phi(E)}{dE}= F \left(\frac{E}{E_{0}}\right)^{-\gamma-\beta \ln \left(\frac{E}{E_{0}}\right)},
\end{equation}
where $F$ is the flux amplitude given in units of $\rm{cm}^{-2}  \, \rm{s}^{-1} \, \rm{TeV}^{-1}$ and $E_0$ is a fixed scale parameter in units of TeV. In this manuscript, the symbol $E$ is designated specifically for the \textit{true} energy, distinguishing it from the \textit{estimated} energy of the gamma ray, which is denoted as $E'$.
Upon conducting a forward folding\footnote{In the forward folding method for SED analysis, an assumed intrinsic spectrum for a source is convolved with the instrument response and adjusted for propagation effects such as the absorption due EBL, thereby 
allowing us to compute the expected numbers of counts for each bin of estimated energy, which are subsequently compared to
the observed data (the On and Off counts in energy) in a maximum-likelihood fit. The spectral parameters that maximize the likelihood function are then found, yielding the best-fitting intrinsic spectrum for the source.}, we obtain the parameters $\gamma$ and $\beta$ for each run, as illustrated in Figure \ref{fig:SED_paramters}. As it is visible from Figure \ref{fig:SED_paramters}, the SED parameters can be reasonably approximated as constant over time: for each run, the values of $\gamma$ are consistent with $\gamma = 2.11 \pm 0.03$ (fitting with a constant model results in $\chi^2/\mathrm{dof} = 1.9$, p-value = 6.4\%), while the $\beta$ values align well with a constant $\beta = 0.119 \pm 0.014$ ( $\chi^2/\mathrm{dof} = 0.4$, p-value = 91\%).

\section{\label{sec:LIV_Analysis}Probing Lorentz invariance violation through a binned likelihood analysis}

\begin{figure}[t]
    \centering
    \includegraphics[width=\textwidth]{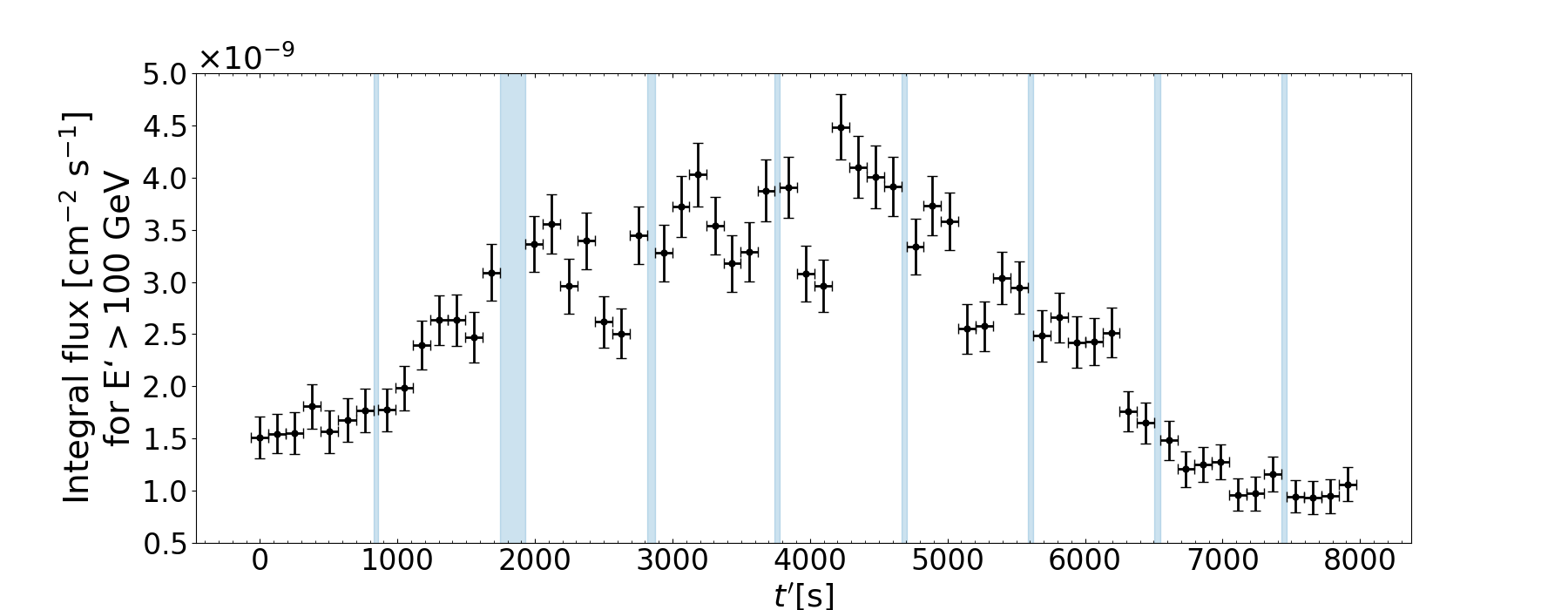} \\
    \caption{
    The plot displays the light curve of Mrk 421 as utilized in our study, with the intervals between observational runs highlighted in light blue. Each time bin has an approximate width of 124 seconds.
    }
    \label{fig:ExcessRate}
\end{figure}

ToF LIV searches are traditionally performed using an unbinned likelihood analysis (see \citep{acciari2020bounds,LimitMAGICMrk501} for instance), introduced in 2009 \cite{ApproachLikeManel}.
In this work, instead of employing an unbinned likelihood analysis, we opted to implement for the first time a binned likelihood analysis. The reason for such a choice relies upon the fact that for an unbinned likelihood
\begin{equation}
    \mathcal{L}(\eta_n)  = 
    \prod_{i=1}^N P( E'_i ,t'_i | \eta_n ),
\end{equation}
one has to provide an expression for computing the probability $P$ of detecting an event $i$ of estimated energy $E'_i$ at the time $t'_i$ (as for the energy, the primed variable denotes the observed one), given the intrinsic temporal and energy distributions of the $N$ events in the dataset. Obtaining certain intrinsic properties poses a considerable challenge, while others can be effortlessly acquired. 
 LIV effects exclusively influence the arrival times of gamma rays without altering their energies. Considering the low likelihood that intrinsic spectral variations might synchronize with LIV effects to produce the constant observed spectral shape, it is reasonable to directly assume that the intrinsic energy distribution aligns with the one detailed in Section \ref{sec:SED}. 
However, the situation is markedly different for the temporal distribution, where LIV influences observed characteristics. This makes the extraction of the intrinsic light curve (LC) a complex task prone to bias. 
In order to surmount these challenges,  other methods which are not based on a likelihood-maximization analysis and do not require defining a LC template can be implemented (see for instance \cite{Limit4GRBFermi} and \cite{amelino2021vacuo}). 
While
various studies \cite{acciari2020bounds,LimitHESSPKS2155Like, LimitMAGICMrk501, CedricPaper,Limit4GRBFermi} have proposed estimating the intrinsic LC by using the low-energy part of the dataset. This part is less affected by LIV effects, given that such effects are proportional to the gamma-ray energy. This approach involves generating a time-based histogram of the low-energy events to estimate the LC template. This histogram is then interpolated using a specified interpolation algorithm or fitted with an analytical function, the form of which can be theoretically justified (see, for example, \citep{acciari2020bounds}). 
This methodology, however, does not incorporate the uncertainties
inherent in the low-energy LC points into the likelihood maximisation analysis for LIV search. The LC, once fitted using low-energy events, is subsequently considered devoid of uncertainties in the maximum likelihood analysis. Essentially, once a LC is fitted, it is often treated as an absolute entity, neglecting its unavoidable intrinsic uncertainties~\footnote{When the used data set spans several orders of magnitude in energy, the much larger number of low-energy photons renders the associated statistical uncertainty
subdominant compared to the systematical one.}. This may often result in biases in the analysis which are difficult to estimate.
The aforementioned difficulty becomes particularly significant when handling data from the flaring state of an astrophysical source, where the time distribution of gamma-ray events often eludes approximation with any analytical function. This is indeed the case for our dataset, as illustrated in Figure \ref{fig:ExcessRate}. The methodology we propose seeks to bridge this gap by devising a mechanism that naturally integrates these uncertainties into the likelihood analysis. The core idea involves segmenting the LC into sufficiently small time bins and replacing the true (unknown) LC function by a function that returns, uniformly within each considered bin, the average flux intensity within each bin over its duration.


 In our LIV study we adopted a binned (in energy and time) likelihood approach:

\begin{equation}
\mathcal{L} = \prod_{i=1}^{N_t} \prod_{j=1}^{N_E} \mathcal{P}(s_{i,j}, b_{i,j} | N_{\text{on},i,j}, N_{\text{off},i,j}),
\label{eq:Poisson}
\end{equation}
where $N_t$ is the number of time bins excluding empty bins (i.e., those in which no data were collected, highlighted in blue in Figure \ref{fig:ExcessRate}), and $N_E$ is the number of energy bins. The quantities  $s_{i,j}$, $N_{\text{on},i,j}$, $b_{i,j}$, and $N_{\text{off},i,j}$ respectively signify the expected signal counts, 
the observed counts in the On region,
the expected background counts, and 
the observed counts in the Off region,
each in the $i$-th time and $j$-th energy bin. Furthermore, $\alpha$ is the ratio of exposure in the On and Off regions, specifically set to $1/3$ for this analysis (the systematic uncertainties on this value are investigated in appendix \ref{Systematics}).
$\mathcal{P}$ is the product of the On and the Off Poisson terms expressed as
\begin{equation}
\mathcal{P}(s,b) = \frac{(s + \alpha b)^{N_{\text{on}}}}{N_{\text{on}}!} e^{-(s + \alpha b)} \frac{b^{N_{\text{off}}}}{N_{\text{off}}!} e^{-b}.
\end{equation}

 Employing the binned likelihood outlined in Equation (\ref{eq:Poisson})  alleviates the issues linked to speculating on the intrinsic light curve.  The comprehensive comparison between our new approach and the conventional unbinned likelihood method will be thoroughly reported in an upcoming dedicated paper.




In Equation (\ref{eq:Poisson}) the expected signal counts $s_{i,j}$ in the $i$-th estimated time bin and $j$-th estimated-energy bin is given by\footnote{In this manuscript, we use the notation $\Delta x_i$ to denote both the bin interval $[x_i, x_{i+1}]$, and the width of the bin, calculated as $x_{i+1} - x_{i}$. This dual usage of $\Delta x_i$ applies coherently throughout the text for simplicity and ease of understanding.}

\begin{equation}
s_{i,j} = \int_{\Delta E'_j} dE' \int_{\Delta t'_i} dt' \int dE \int dt \;  \frac{d\Phi(E, t)}{dE} \; B(E) \; A(E,t') \;  G(E' | E, t') \; T( t'| E, t)
\label{Eq:exp_signal_formula_1}
\end{equation}
where:
\begin{itemize}


    
    \item $\frac{d\Phi(E,t)}{dE} $ is the intrinsic differential flux per unit energy.

    \item $B(E)$  represents the photon-survival probability due to the absorption by the EBL, which we model following Ref. \cite{EBLDominguez}.

    \item $A(E,t') $ is the effective
    collection area of the detector at time $t'$, and $G(E' | E, t')$ is the probability density for a gamma ray with true energy $E$ detected at time $t'$ to be assigned an estimated energy $E'$. Both $A$ and $G$ are obtained from Monte Carlo simulations, while for times between runs (those highlighted in blue in Figure \ref{fig:ExcessRate}), the collection area is fixed to zero, $A=0$.

    \item  $T( t'| E, t)$ is the probability density  for a gamma ray with true energy $E$ and arrival time in case of no LIV effect, $t$, to arrive at the time $t'$, which can be written as  
    \begin{equation}
     T( t'| E, t) dt' = \delta ( t' - t - \eta_n E^n)dt',
\end{equation}
where $\delta$ is the Dirac delta function.

\end{itemize}

The integrals in Eq. (\ref{Eq:exp_signal_formula_1}) over $E $ and $t$ perform the convolution
of the gamma-ray spectrum with the instrumental response and introduce the LIV effects in the observed flux, respectively. Those over $E'$ and $t'$ compute the expected number of gamma rays within the i-th arrival time bin $\Delta t'_i$ and the j-th estimated energy bin $\Delta E'_j$. 

Integrating over $t'$, Eq. (\ref{Eq:exp_signal_formula_1}) becomes
\begin{equation}
s_{i,j} = \int_{\Delta E'_j} dE'  \int dE \int dt \;  \frac{d\Phi(E, t)}{dE} \; B(E) \; A_i(E) \;  G_i(E' | E ) I_i(t+\eta_n E^n)
\label{Eq:exp_signal_formula_2}
\end{equation}
with $ A_i(E)$ and $G_i(E' | E ) $ the time-average in $\Delta t'_i$ of $G(E' | E, t')$ and $A(E,t') $, respectively, and $I_i(t) $ the ``indicator''  or  ``characteristic'' function:
\begin{equation}
I_i(t) = 
\begin{cases}
1  & \; \rm{if} \; t \in \Delta t'_i 
\\
 0 &\;  \rm{otherwise}.
\end{cases}
\end{equation}
At this point, one can approximate 
\footnote{  This approximation holds under the condition that the bin sizes, $\Delta t_k$, are selected to be smaller than the analysis's sensitivity to detecting or ruling out specific values of the spectral-lag parameter, $\eta$. Thus to justify this approximation, we check that varying the assumed temporal distribution of photons within each time bin, does not affect significantly the outcomes of the analysis, i.e., the change is within the systematic uncertainties discussed in appendix \ref{Systematics}. }
the evolution of the flux in time as follows:

\begin{equation}
    \frac{d\Phi(E,t)}{dE} \approx   \frac{d \Phi_k(E)}{dE} \quad \text{for} \ t \in \Delta t_k \quad (k = 1, \ldots, N_{\text{bin}}),
\end{equation}
with $\Phi_k$ the time-independent flux in the $k$-th bin, and  integrate the variable $t$ in Eq. (\ref{Eq:exp_signal_formula_2}), obtaining

\begin{equation}
s_{i,j} = \sum_{k=1}^{N_{\text{bin}}} \int_{\Delta E'_j} dE'  \int dE \; \frac{d \Phi_{k} (E)}{d E }  \; B(E)  \; A_i(E) \;  G_i(E' | E ) \; \Delta t_{i,k}(\eta_n , E),
\label{Eq:exp_signal_formula_3}
\end{equation}
in which we have defined
\begin{equation}
    \Delta t_{i,k}(\eta_n , E) \equiv \int_{\Delta t_k} dt \;  I_i(  t + \eta_n E^n)  .
    \label{Eq:DeltaT_matrix}
\end{equation}
 $\Delta t_{i,k}(\eta_n , E)$ is an energy-dependent matrix that considers the intrinsic flux contribution from the $k$-th time bin $\Delta t_k$ to the $i$-th time bin $\Delta t'_i$, factoring in the LIV-induced delays. It encapsulates all the LIV-related information and functions similarly to a migration matrix, where the migration occurs between time bins. Once having defined a binning in $t$ and $t'$, for a given true energy $E$, this matrix can be easily computed from Equation (\ref{Eq:DeltaT_matrix}).
For example, in scenarios where no LIV effect is present (i.e., $\eta_n = 0$), and assuming that the binning for $t$ and $t'$ is identical, the $\Delta t_{i,k}$ becomes proportional to the identity matrix:
\begin{equation}
\Delta t_{i,k}(0,  E)
= 
\begin{pmatrix}
\Delta t_0 & 0 &   &  \dots \\
 0 & \Delta t_1  & 0 &  \dots\\
 &  & \vdots& \\
 & \dots & 0 & \Delta t_{N_t} & 
\end{pmatrix}.
\label{Eq:Matrix_form_NO_LIV}
\end{equation}
If we introduce a LIV-induced time delay $d = \eta_n E^n$ smaller than the bin width, Equation \ref{Eq:Matrix_form_NO_LIV} can be written as follows:
\begin{equation}
\Delta t_{i,k}(\eta_n,  E)
= 
\begin{pmatrix}
 \Delta t_0 - d & 0 & \dots &  & & & &   \\
 d & \Delta t_1 - d  & 0 & \dots & & & & \\
  0 & d& \Delta t_2 - d  & 0 &   \dots & & & \\
  &  & &  \vdots & &  &  &\\
&  & &  & \dots & \;  0   & \quad d & \quad \Delta t_{N_t} - d
\end{pmatrix}.
\label{Eq:Matrix_form_LIV}
\end{equation}

 We have already shown in Section \ref{sec:SED} that the gamma-ray flux can be well described by a log-parabola, therefore:
 \begin{equation}
    \frac{d\Phi_k(E,t)}{dE} = F_k  \left( \frac{E}{E_0} \right)^{ -\gamma_k - \beta_k \ln(E/E_0)} \quad \text{for} \ t \in \Delta t_k, 
    \label{eq:flux_energy}
\end{equation}
with $F_k$ the flux amplitude in the $k$-th time bin, while, as discussed in Section \ref{sec:SED}, the other parameters can be assumed to not change in time, i.e., $\gamma_k = \gamma$ and $\beta_k = \beta$.

From Equation \ref{Eq:exp_signal_formula_3}, we can define in a more compact form the expected signal counts as:

\begin{equation}
s_{i,j} = \sum_{k=1}^{N_{\text{bin}}} M_{i,j}^k(\eta_n) F_k,
\label{Eq:s_ij_tensor}
\end{equation}
where

\begin{equation}
M_{i,j}^k(\eta_n) = \int_{\Delta E'_j} dE'  \int dE \; \frac{d \tilde{\Phi}_{k} (E)}{d E }  \; B(E)  \; A_i(E) \;  G_i(E' | E ) \; \Delta t_{i,k}(\eta_n , E),
\end{equation}
with $\tilde{\Phi}_{k} \equiv \Phi_{k}/F_k$ a constant for all bins.

Taking the logarithm of Equation \ref{eq:Poisson} and using the expression in  Equation \ref{Eq:s_ij_tensor},  the log-likelihood can be expressed as (ignoring constant terms)

\begin{equation}
-2 \ln \mathcal{L} = 2 \sum_{i=1}^{N_t} \sum_{j=1}^{N_E} \left( M_{i,j}^k F_k - N_{\text{on},i,j} \ln(M_{i,j}^k F_k + \alpha b_{i,j}) + (\alpha + 1) b_{i,j} - N_{\text{off},i,j} \ln b_{i,j} \right),
\label{Eq:log-likelihood}
\end{equation}
where for the sake of clarity, the multiplication between the matrix $M_{i,j}^k $ and vector $F_k$ is written adopting the Einstein notation.\\ 
The log-likelihood ratio, used as the test statistic for the search of LIV effects in this analysis, is given by:

\begin{equation}
-2 \Delta \ln \mathcal{L}(\eta_n) = -2 \ln \mathcal{L}(\eta_n; \mathbf{\tilde{F}}, \mathbf{\tilde{b}}, \tilde{\gamma}, \tilde{\beta}) + 2 \ln \tilde{\mathcal{L}},
\label{Eq:likelihood_ratio}
\end{equation}
where $\tilde{\mathcal{L}}$ signifies the maximum value of the likelihood, while for all nuisance parameters in the analysis, the ``tilde'' on top of the variable represents the values of that variable that maximize the likelihood for a given $\eta_n$~\footnote{It is important to note that during the likelihood profiling process, the time-dependent parameters in our model cannot be constrained by our observations, for specific values of $\eta_n$ and for the values of $t^\prime$ corresponding to the intervals between runs, since no data are collected by the telescopes.}.
According to the Wilks' theorem \citep{Wilks:1938} (we discuss its applicability in our case in appendix \ref{coverage}) the profile likelihood ratio (\ref{Eq:likelihood_ratio}) evaluated at the true value of $\eta_n$ follows a $\chi^2$-distribution with one degree of freedom (dof), which allows a straightforward way of obtaining the $95\%$ ULs $\eta_n^{95}$ on the spectral lag parameter:
\begin{equation}
    -2 \Delta \ln \mathcal{L}(\eta_n^{95}) = 3.84.
    \label{eq:likelihood_UL}
\end{equation}

\section{\label{sec:Results}Results}

The computation of the likelihood defined in Equation~(\ref{Eq:log-likelihood}) requires defining an estimated energy $E'$ and time $t'$ binning. The selection of the number of bins is guided by two balancing factors:

\begin{itemize}
\item The minimum detectable LIV-induced time delay is constrained by the size of the time bins, as the information contained within each bin gets averaged out. Consequently, an excessively small number of time bins will unnecessarily curtail the sensitivity of our analysis.
\item An excessive number of bins can lead to a computationally intensive likelihood profiling process, as the number of nuisance parameters (see Equation \ref{Eq:likelihood_ratio}) is directly proportional to the number of bins: while it is possible to analytically determine the expected background counts $b$ that maximize the likelihood when all other parameters are set, this is not the case for the SED parameters $\gamma$, $\beta$ and the flux amplitude per time bin $F_k$ (see Equation \ref{Eq:log-likelihood}). Consequently, determining the values of $F_k$, $\gamma$ and $\beta$ that optimize the likelihood for a given spectral lag $\eta_n$ necessitates the numerical implementation of an optimization algorithm. However, as the number of time bins increases, this search can become computationally prohibitive. 
\end{itemize}

Given the considerations stated above, we selected a binning arrangement for our dataset comprising 70 bins in time (on a linear scale from $T_0$ up to $T_0 + 8159$ s) and 10 bins in energy (of equal size on a logarithmic scale from 100 GeV up to 10 TeV). 
In appendix \ref{Systematics} we will discuss how the change of binning affects the result of the likelihood analysis. 

    

\begin{figure}[t]
    \centering
    \includegraphics[width=0.49\textwidth]{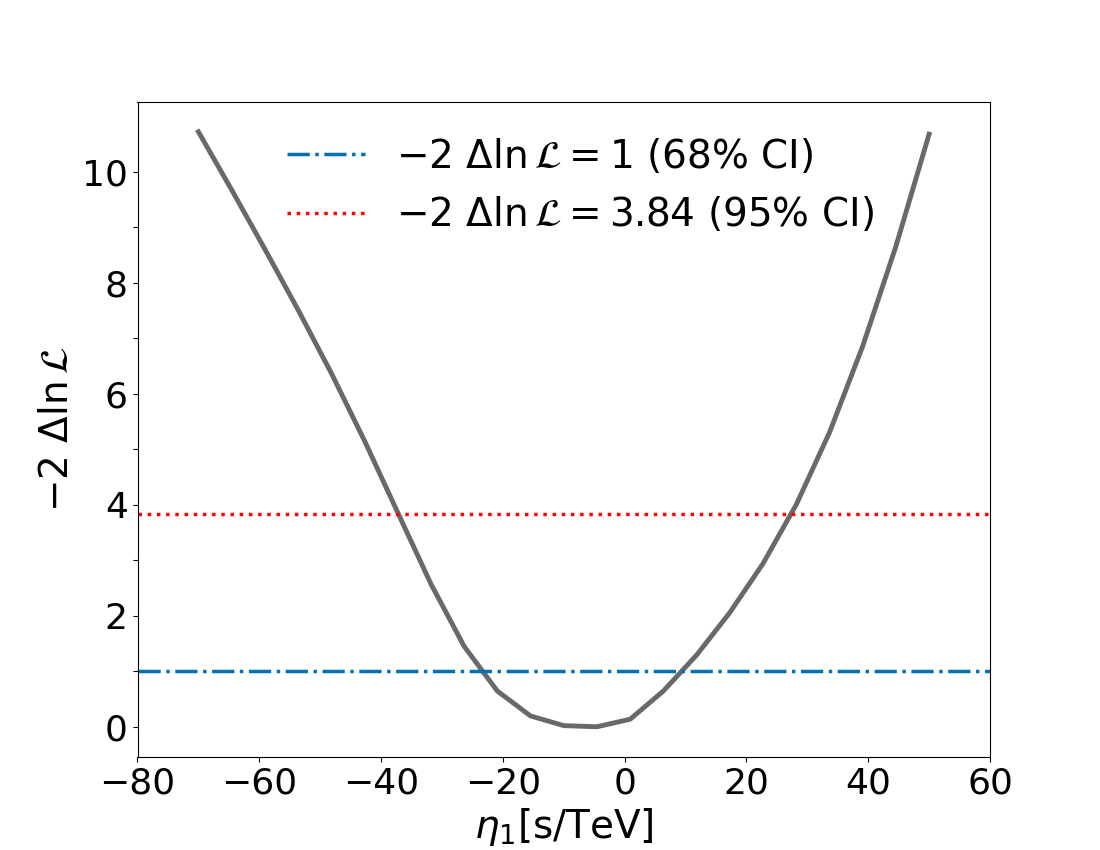}
    \includegraphics[width=0.49\textwidth]{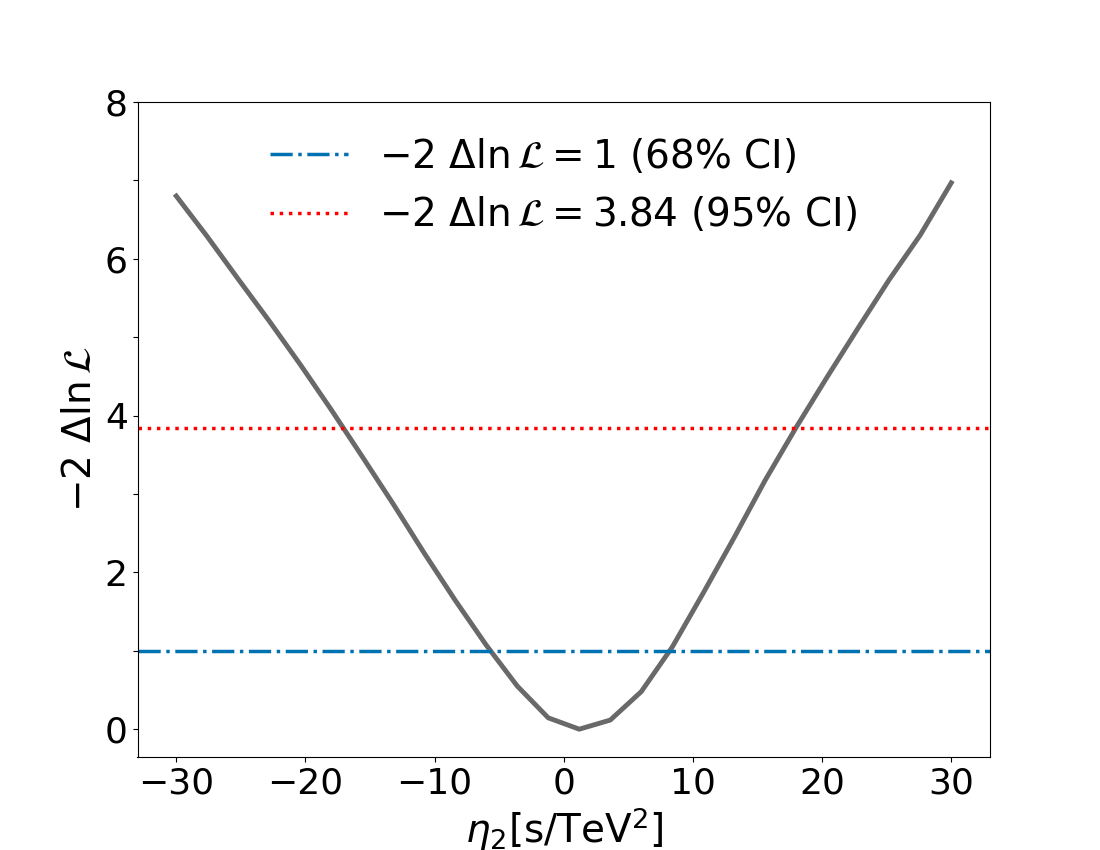}
    \caption{Log-likelihood ratio $-2 \Delta \ln \mathcal{L}(\eta_n) $ defined in Equation (\ref{Eq:likelihood_ratio}) obtained from the Mrk\,421 data described in Section \ref{sec:DataSet}. The horizontal lines in blue and red are for $-2 \Delta \ln \mathcal{L} = $ 1 and 3.84, corresponding to a $68\%$ and $95\%$ CI, respectively. Left: Linear scenario. Right: quadratic scenario.} 
    \label{fig:Likelihood_Profile}
\end{figure}

As discussed in Section~\ref{sec:SED}, we adopted a log-parabola for the energy spectrum, where the global (time-independent) parameters $\gamma$ and $\beta$ were treated as nuisance parameters~\footnote{Upon profiling the likelihood over the nuisance parameters (i.e., determining the values of these parameters that maximize the likelihood), we derived values of $\gamma = 2.115$ and $\beta = 0.124$, which vary by less than $1\%$ when changing the parameter $\eta_n$. These findings are consistent, within their respective uncertainties, with the values presented in Section \ref{sec:SED}.}. The profile likelihood ratio (see Equation \ref{Eq:likelihood_ratio}) as a function of the spectral lag $\eta_n$, for the linear and quadratic case, is reported in Figure~\ref{fig:Likelihood_Profile}.

The null hypothesis of no spectral lag
is compatible with the observation as the value $\eta_n = 0 $ lies in the $68\%$ confidence interval (CI) defined by the interval in which $-2 \Delta \ln \mathcal{L} \leq 1$ (see Equation \ref{Eq:likelihood_ratio}).

From Equation (\ref{eq:likelihood_UL}), we derive the following $95\%$ CL limits on the spectral lag $\eta_1$ for the linear case:

\begin{equation}
\eta_1^{95} = - 38 \; \text{s/TeV}, \; \; + 31  \; \text{s/TeV} .
\end{equation}
Considering the linear case ($n=1$) in Equation (\ref{eq:spectral_lag}), where $\kappa_1(z_\mathrm{s}=0.03) = 1.33 \times 10^{16} $ seconds is obtained from Equation (\ref{eq:distanceToFPiran}), we can establish LLs on the quantum-gravity energy scale $E_\mathrm{QG,1}$ for both superluminal and subluminal cases. At a CL of $95\%$, these LLs are $3.5 \times 10^{17}$ GeV and $4.8 \times 10^{17}$ GeV, respectively.

In a similar way, the quadratic case provides us with:

\begin{equation}
\eta_2^{95} = - 16 \; \text{s/TeV}^2, \; \; + 16 \; \text{s/TeV}^2 .
\end{equation}
Considering $\kappa_2(z_\mathrm{s}=0.03) = 1.35 \times 10^{16} $ seconds in Equation (\ref{eq:spectral_lag}), these values can be translated at a CL of $95\%$ into LLs of $3.6 \times 10^{10}$ GeV and $3.5 \times 10^{10}$ GeV for the quantum-gravity energy scale $E_\mathrm{QG,2}$ in the superluminal and subluminal scenarios, respectively.

\section{\label{sec:Discussion}Discussion and conclusions}

In this work, we establish new constraints on Lorentz invariance violation by the analysis of the extraordinary flare of Mrk\,421 observed with the MAGIC telescopes on the night of April 25 to 26, 2014. Our findings for both the linear and quadratic cases, as detailed in Table~\ref{tab:QGlimits}, demonstrate a modest change in the Quantum Gravity (QG) energy scale estimate when systematic uncertainties, discussed in detail in appendix ~\ref{Systematics}, are considered. The derived LLs on the QG energy scale across all investigated scenarios — subluminal and superluminal for both linear and quadratic cases —  are comparable to the previous constraints obtained using 
AGN data.

\begin{table}
    \centering
\begin{tabular}{ |l|l|l| }
  \hline
   \multicolumn{3}{|c|}{\textbf{Obtained limits}} \\\hline
  Case & \textbf{No systematic} & \textbf{Including systematic}\\
    & \textbf{ uncertainties} & \textbf{ uncertainties}\\\hline
    \multicolumn{3}{|c|}{\textbf{Linear scenario: $E_\mathrm{QG,1}/ \mathrm{GeV} $}} \\ \hline
  superluminal & $3.5\times10^{17}$ &  $2.7\times10^{17}$\\
  subluminal & $4.8\times10^{17}$ & $3.6\times10^{17}$ \\ \hline
  \multicolumn{3}{|c|}{\textbf{Quadratic scenario: $E_\mathrm{QG,2}/ \mathrm{GeV} $}} \\ \hline
  superluminal & $3.6\times10^{10}$ & $2.6\times10^{10}$ \\
  subluminal & $3.5 \times10^{10}$ & $2.5\times10^{10}$ \\
    \hline
\end{tabular}
\caption{95\% LLs on the QG energy scale without and with the systematic uncertainties.}
\label{tab:QGlimits}
\end{table}
%
%
%
%
%
%
%
%
%
While the limits on the QG energy scale determined in this study are lower than those found in previous analyses, it is noteworthy to highlight that our methodology diverges from earlier approaches. The method employed here incorporates all relevant uncertainties in calculating these limits, which cannot be neglected during episodes such as the April 2014 Mrk 421 flare, which is marked by a complex temporal distributions.

\section*{Author contributions}
 G. D'Amico led the statistical analysis and interpretation of the results. L. Nogués Marcén led the initial data and LIV analysis. J. Strišković led the data analysis and assisted in the interpretation of the results and paper drafting. T. Terzić assisted in the data analysis and interpretation of the results.  The rest of the authors have contributed in one or several of the following ways: design, construction, maintenance and operation of the instrument(s); preparation and/or evaluation of the observation proposals; data acquisition, processing, calibration and/or reduction; production of analysis tools and/or related Monte Carlo simulations; discussion and
approval of the contents of the draft.

\acknowledgments
G.D'A.’s work on this project was supported by the Research Council of Norway, project number 301718.   G.D'A. would like to thank the Institut de F\'{\i}sica d'Altes Energies (IFAE) for hosting him during the analysis and production of the manuscript.
We would like to thank the Instituto de Astrof\'{\i}sica de Canarias for the excellent working conditions at the Observatorio del Roque de los Muchachos in La Palma. The financial support of the German BMBF, MPG and HGF; the Italian INFN and INAF; the Swiss National Fund SNF; the grants PID2019-104114RB-C31, PID2019-104114RB-C32, PID2019-104114RB-C33, PID2019-105510GB-C31, PID2019-107847RB-C41, PID2019-107847RB-C42, PID2019-107847RB-C44, PID2019-107988GB-C22, PID2022-136828NB-C41, PID2022-137810NB-C22, PID2022-138172NB-C41, PID2022-138172NB-C42, PID2022-138172NB-C43, PID2022-139117NB-C41, PID2022-139117NB-C42, PID2022-139117NB-C43, PID2022-139117NB-C44 funded by the Spanish MCIN/AEI/ 10.13039/ 501100011033 and “ERDF A way of making Europe”; the Indian Department of Atomic Energy; the Japanese ICRR, the University of Tokyo, JSPS, and MEXT; the Bulgarian Ministry of Education and Science, National RI Roadmap Project DO1-400/18.12.2020 and the Academy of Finland grant nr. 320045 is gratefully acknowledged. This work was also been supported by Centros de Excelencia ``Severo Ochoa'' y Unidades ``Mar\'{\i}a de Maeztu'' program of the Spanish MCIN/AEI/ 10.13039/501100011033 (CEX2019-000920-S, CEX2019-000918-M, CEX2021-001131-S) and by the CERCA institution and grants 2021SGR00426 and 2021SGR00773 of the Generalitat de Catalunya; by the Croatian Science Foundation (HrZZ) Project IP-2022-10-4595 and the University of Rijeka Project uniri-prirod-18-48; by the Deutsche Forschungsgemeinschaft (SFB1491) and by the Lamarr-Institute for Machine Learning and Artificial Intelligence; by the Polish Ministry Of Education and Science grant No. 2021/WK/08; and by the Brazilian MCTIC, CNPq and FAPERJ.


\bibliographystyle{JHEP}
\bibliography{LIV_Mrk421-2014_forJCAP}



\appendix
\section{\label{sec:Systematics}Coverage and systematic uncertainties}

\subsection{\label{coverage}Coverage}

According to Wilks' theorem \citep{Wilks:1938}, the number of dof is equivalent to the number of independent parameters 
(in our analysis, there is only one: the spectral lag parameter). The likelihood ratio statistic, as outlined in Equation (\ref{Eq:likelihood_ratio}), is expected to follow, if the null hypothesis ($\eta=0$) is true, a chi-squared distribution. 
To validate the applicability of Wilks' theorem and ensure the coverage accuracy assumed in the likelihood ratio of Equation (\ref{Eq:likelihood_ratio}), we compute the log-likelihood ratio from Equation (\ref{Eq:log-likelihood}) with $\eta_n = 0$ (representing the null hypothesis). 
This computation is performed on 100 simulated datasets. Each set is generated from the measured data, first through shuffling of event arrival times and then using a bootstrap resampling technique 
(the same method was used to extract the CI in \cite{acciari2020bounds}).
Shuffling involves randomly reassigning measured arrival times to different observed events, effectively erasing any energy-time correlation present in the data, including any potential LIV effects. However, this operation does not alter the overall spectral and temporal distributions of the signal.
On the other hand, the bootstrap technique generates samples of equal size by randomly selecting events (with repetitions allowed) from the shuffled dataset. This process permits the measured spectral and temporal distributions to naturally fluctuate within their statistical uncertainties.

\begin{figure}[h]
    \centering
    \includegraphics[width=0.8\textwidth]{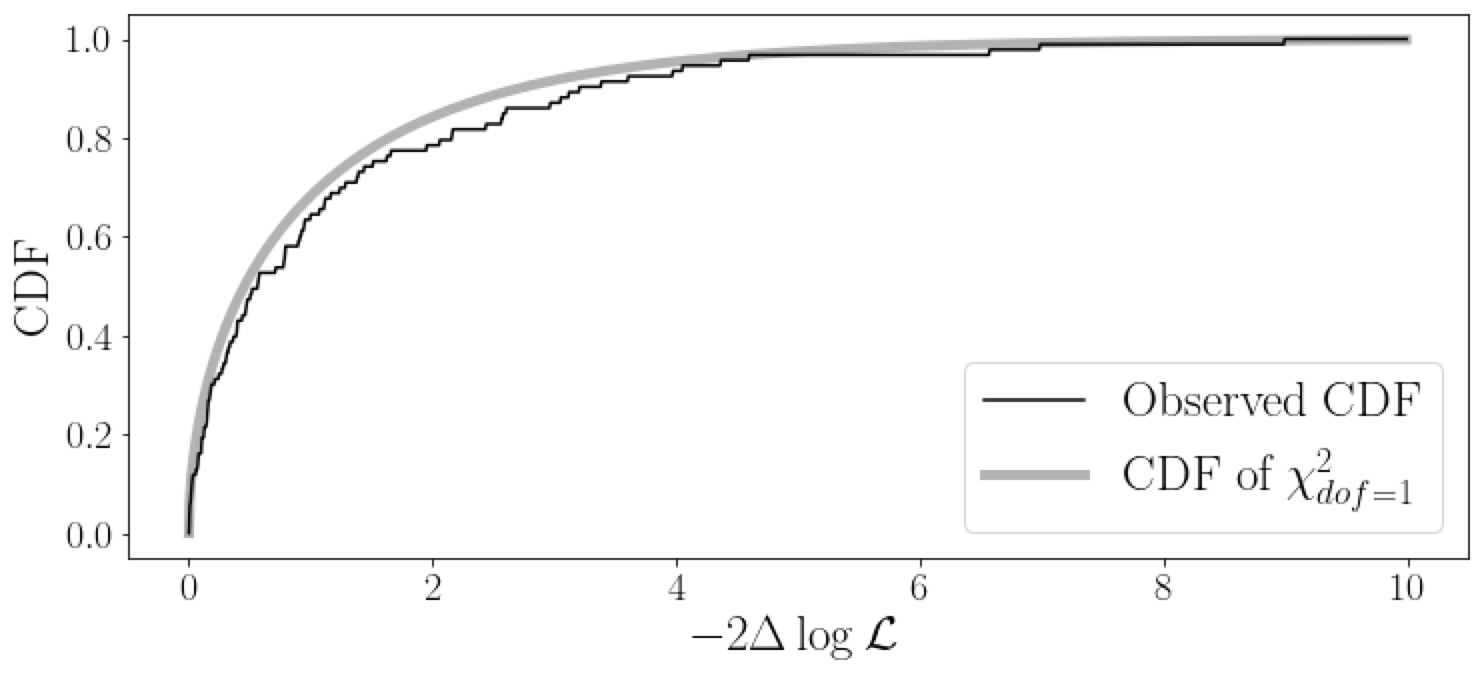}

    \caption{Cumulative distribution function (CDF) of the likelihood ratio obtained from Monte Carlo simulations assuming the null hypothesis to be true (black line). This is compared with the CDF of a $\chi^2$ distribution with one dof (grey line), showcasing good agreement between the two. The alignment justifies the application of Wilks' theorem in this analysis.} 
    \label{fig:CDF_TS}
\end{figure}
The results of these Monte Carlo (MC) simulations are illustrated in Figure \ref{fig:CDF_TS}, where the cumulative distribution function (CDF) of the likelihood ratio is compared with the CDF of a $\chi^2$ distribution with one dof. As seen in Figure \ref{fig:CDF_TS}, the likelihood ratio aligns well with the $\chi^2$ distribution hypothesis, thereby validating~\footnote{
A proper validation would require conducting this test for non-zero values of $\eta_n$.
However, we would need to produce Monte Carlo simulations with a spectral lag, which is not feasible as it demands precise knowledge of the true energy of gamma rays and their exact temporal distributions, information that is not available.} the use of Wilks' theorem in this analysis: the application of the Kolmogorov-Smirnov test results in a p-value of $40.1\%$, indicating a good level of agreement between the simulated log-likelihood ratios and the expected $\chi^2$ distribution with one dof.
Out of the 100 computed likelihood ratios, only 7 exceed the value of 3.84. Defining $x$ as the variable representing the number of times the value of the likelihood ratio is above a given threshold (3.84 according to Wilks' theorem), to achieve the desired value of $x=5$, we must increase the threshold from 3.84 to 4.16. Applying this adjusted threshold of 4.16 in Equation \ref{eq:likelihood_UL} would result in less than a $5\%$ change in the UL of $\eta_n$. The small size of our sample, comprising only 100 simulations, was constrained by the substantial computational power required to compute the log-likelihood ratio for each simulated sample.

\subsection{\label{Systematics}Systematic uncertainties}

This section covers the various systematic uncertainties that could potentially impact the analysis, as detailed in Table \ref{tab:Systematics_table} and outlined below:

\begin{table}[h]
\centering
\begin{tabular}{ |l|l|l| }
  \hline
   \multicolumn{3}{|c|}{\textbf{Study of systematic uncertainties}} \\\hline
  \textbf{Systematic effect} & \textbf{Size($E_\mathrm{QG,1}$)} & \textbf{Size($E_\mathrm{QG,2}$)} \\  \hline
  Number of bins in time & < 14\% & < 16\% \\
  Number of bins in energy & < 11\% & < 17\% \\
  Energy scale & $\sim$ 15\% & $\sim$ 15\% \\
  Background normalization & <0.1\% & <0.1\% \\
  Cosmological model & <4\% & <5\% \\ \hline
  Total & $\lesssim$ 24 \% & $\lesssim$ 28 \% \\
  \hline
\end{tabular}
\caption{List of the systematic uncertainties examined in our study and their impact on determining the LL of the QG energy scale. For uncertainties that fluctuate based on parameter choices, such as the number of bins in time, the most conservative estimate has been reported. The total systematic uncertainty was calculated by taking the square root of the sum of the squares of the individual uncertainties.}
\label{tab:Systematics_table}
\end{table}

\begin{itemize}
    \item {\bf Number of Bins:}
    
    To assess the influence of the chosen number of bins in time and energy in our analysis, we replicate the analysis with varying bin number. Initially, we alter the number of time bins while keeping the energy bins fixed.  From this, we deduce that changing the number of time bins can affect the LL up to $14\%$ in the linear case and $16\%$ in the quadratic case.
Similarly, we varied the number of energy bins from 5 to 15, while keeping the number of time bins constant, and computed the LL on the QG energy scale. We found that variations in the number of energy bins could affect the LL by up to $11\%$ in the linear case and $17\%$ in the quadratic case.

\item {\bf Energy scale: }

    The calibration of the energy scale is one of the most significant systematic effects impacting observations with IACTs. For the MAGIC telescopes, uncertainties on the energy scale are estimated to be around $15\%$ ~\citep{MAGICUpgradeSoftware}. To quantify this effect, we artificially increased or decreased the variable $E$ in Equation (\ref{eq:delay}) by $\pm 15\%$ and repeated the likelihood analysis. 
    For both the linear and quadratic scenarios, we estimated the impact of this adjustment to be $\sim 15\%$.

\item {\bf Background normalization: }

Variations in the photomultiplier tube (PMT) response and Night Sky Background (NSB) across the telescopes' field of view, together with the natural inhomogeneity resulting from stereoscopic observation using two telescopes, lead to asymmetries in camera acceptance. These inhomogeneities can be mitigated by the wobble observation mode, in which the positions for source and background estimation region in the camera are alternated. 
While the observational inhomogeneities of the MAGIC telescopes introduce an estimated uncertainty of less than $1\%$ ~\citep{MAGICUpgradeSoftware}, this effect is virtually negligible in the context of a strong source such as the Mrk\,421 flare analyzed in this study. Nevertheless, to account for this potential source of error, we adjust the $\alpha$ value by $\pm 1\%$ and recompute the likelihood to derive new LLs on the QG energy scale. Our findings indicate that this systematic effect may alter the limits by a marginal $0.1\%$, reinforcing its negligible impact on our analysis.

    \item {\bf Cosmological model: }

    In our analysis we used the $\Lambda$CDM model with a nominal Hubble constant ($H_0$) value of 70\,km/s/Mpc. However, recent cosmological measurements suggest a range of values, contributing to the ongoing debate often referred to as the ``Hubble tension''. The Planck Collaboration in 2021, using data from the cosmic microwave background radiation  reported an estimate of $H_0 = 67.4 \pm 0.5$\,km/s/Mpc \cite{aghanim2020planck}. Conversely, a study in 2019 \cite{riess2019large}, employing a different approach based on the distance of Cepheid variable stars, suggest a higher value of $H_0 = 74.0 \pm 1.4$\,km/s/Mpc.
    To gauge the influence of these diverging $H_0$ values on our results, we conducted additional analyses. With the lower $H_0$ value from the Planck Collaboration, the LL on the QG energy scale increases by $ 4\%$ and  $ 2\%$ for the linear and quadratic case, respectively. On the other hand, employing the higher $H_0$ value suggested in \citep{riess2019large} results in a decrease of $5\%$ and $3\%$ in the estimated LL on the QG energy scale, for the linear and quadratic case, respectively. Given the relatively short distance to Mrk\,421, it might seem suitable to apply the Hubble constant ($H_0$) value derived from local sources. However, utilizing a generic value of 70 km/s/Mpc enables a straightforward comparison with earlier Lorentz symmetry tests, at least until the current Hubble tension is resolved. Considering the marginal systematic uncertainty induced by this selection, this choice seems as a prudent one.

\end{itemize}

\end{document}

%% file: commands.tex
\newacronym{gr}{GR}{General Relativity}
\newacronym{sr}{SR}{Special Relativity}
\newacronym{qm}{QM}{Quantum Mechanics}
\newacronym{qft}{QFT}{Quantum Field Theory}
\newacronym{eft}{EFT}{Effective Field Theory}
\newacronym{qt}{QT}{Quantum Theory}
\newacronym{lqg}{LQG}{Loop Quantum Gravity}
\newacronym{st}{ST}{String Theory}
\newacronym{liv}{LIV}{Lorentz Invariance Violation}
\newacronym{tof}{ToF}{Time-of-Flight}
\newacronym{qg}{QG}{Quantum Gravity}
\newacronym{ml}{ML}{Maximum Likelihood}
\newacronym{pdf}{PDF}{Probability distribution function}
\newacronym{sme}{SME}{Standard Model Extension}
\newacronym{dsr}{DSR}{Double Special Relativity}
\newacronym{grb}{GRB}{Gamma-Ray Burst}
\newacronym{agn}{AGN}{Active Galactic Nucleus}
\newacronym{magic}{MAGIC}{Major Atmospheric Gamma-ray Imaging Cherenkov}
\newacronym{iact}{IACT}{Imaging Atmospheric Cherenkov Telescope}
\newacronym{ebl}{EBL}{Extragalactic Background Light}
\newacronym{mad}{MAD}{Median Absolute Deviation}

%% file: opted-in_collaborators_jcap_poll_2177_date_2023-11-29.tex
\author [1] {S.~Abe}
\author [2] {J.~Abhir}
\author [3] {A.~Abhishek}
\author [4] {V.~A.~Acciari}
\author [5] {A.~Aguasca-Cabot}
\author [6] {I.~Agudo}
\author [7] {T.~Aniello}
\author [8,43] {S.~Ansoldi}
\author [7] {L.~A.~Antonelli}
\author [9] {A.~Arbet Engels}
% ...

\affiliation [1] {Japanese MAGIC Group: Institute for Cosmic Ray Research (ICRR), The University of Tokyo, Kashiwa, 277-8582 Chiba, Japan}
\affiliation [2] {ETH Z\"urich, CH-8093 Z\"urich, Switzerland}
\affiliation [3] {Universit\`a di Siena and INFN Pisa, I-53100 Siena, Italy}
\affiliation [4] {Instituto de Astrof\'isica de Canarias and Dpto. de  Astrof\'isica, Universidad de La Laguna, E-38200, La Laguna, Tenerife, Spain}
\affiliation [5] {Universitat de Barcelona, ICCUB, IEEC-UB, E-08028 Barcelona, Spain}
\affiliation [6] {Instituto de Astrof\'isica de Andaluc\'ia-CSIC, Glorieta de la Astronom\'ia s/n, 18008, Granada, Spain}
\affiliation [7] {National Institute for Astrophysics (INAF), I-00136 Rome, Italy}
\affiliation [8] {Universit\`a di Udine and INFN Trieste, I-33100 Udine, Italy}
\affiliation [9] {Max-Planck-Institut f\"ur Physik, D-85748 Garching, Germany}
% ...
\affiliation [43] {also at International Center for Relativistic Astrophysics (ICRA), Rome, Italy}
% ...
